\documentclass[10pt,conference]{IEEEtran}
\IEEEoverridecommandlockouts
\usepackage{cite}
\usepackage{amsmath,amssymb,amsfonts,physics}
\usepackage{algorithmic}
\usepackage{graphicx}
\usepackage{textcomp}
\usepackage{xcolor}

\newcommand{\cnot}{\textsc{cnot}}
\newcommand{\cphase}{\textsc{cphase}}

\def\BibTeX{{\rm B\kern-.05em{\sc i\kern-.025em b}\kern-.08em
    T\kern-.1667em\lower.7ex\hbox{E}\kern-.125emX}}
\begin{document}

\title{Efficient Quantum Gate Discovery \\ with Optimal Control}

\author{
\IEEEauthorblockN{Paul Kairys and Travis S.~Humble}
\IEEEauthorblockA{
\textit{Bredesen Center},
\textit{University of Tennessee}\\
Knoxville, Tennessee, United States of America}
\IEEEauthorblockA{\textit{Quantum Science Center},
\textit{Oak Ridge National Laboratory}\\
Oak Ridge, Tennessee, United States of America}

\thanks{This manuscript has been authored by UT-Battelle, LLC, under contract DE-AC05-00OR22725 with the US Department of Energy (DOE). The US government retains and the publisher, by accepting the article for publication, acknowledges that the US government retains a nonexclusive, paid-up, irrevocable, worldwide license to publish or reproduce the published form of this manuscript, or allow others to do so, for US government purposes. DOE will provide public access to these results of federally sponsored research in accordance with the DOE Public Access Plan (http://energy.gov/downloads/doe-public-access-plan).}

}

\maketitle

\begin{abstract}
Optimal control theory provides a framework for numerical discovery of device controls that implement quantum logic gates, but common objective functions used for optimization often assign arbitrarily high costs to otherwise useful controls. We propose a framework for designing objective functions that permit novel gate designs such as echo pulses or locally-equivalent gates. We use numerical simulations to demonstrate the efficacy of the new objective functions by designing microwave-only pulses that act as entangling gates for superconducting transmon architectures. We observe that the proposed objective functions lead to higher fidelity controls in fewer optimization iterations than obtainable by traditional objective functions.
\end{abstract}


\section{Introduction}
Within the circuit model of quantum computing, gates play a fundamental role in the demonstration of quantum algorithms and quantum programs. Improving the fidelity of quantum gates is a necessary step to realize a quantum advantage in the noisy intermediate-scale quantum (NISQ) regime \cite{preskill2018quantum} and achieve fault-tolerance in future quantum devices \cite{steane1999efficient}. However, most quantum logic gates implemented today suffer from noise and errors that limit the observed fidelity \cite{humble2019quantum}, and it will be essential to discover efficient quantum processes that realize these quantum gates with high fidelity.
\par
One route to discover quantum gates is through the use of numerical quantum optimal control\cite{palao_quantum_2002,palao2003optimal,werschnik_quantum_2007,glaser_training_2015, kirchhoff_optimized_2018,machnes_tunable_2018}. Within the optimal control paradigm, a model of the quantum device is used to iteratively refine the applied controls that generate a desired quantum evolution. Typically, these controls are determined by optimization of a objective or cost function that assigns a value to each quantum evolution. For purposes of quantum gate discovery, it is therefore critical to assign an operationally significant cost to each quantum evolution. 
\par
In this work we consider the task of constructing a universal two-qubit quantum gate via optimal control. It is a well-known result that nearly any two-qubit unitary operator is universal for quantum computing \cite{lloyd1995almost}, therefore it is desirable to have an optimal control objective function which will assign equivalent costs to universal two-qubit quantum operations. Common objective functions are incapable of realizing this goal, motivating a new approach to objective function design.
\par
We propose a framework to define objective functions for application-specific gate discovery tasks. We demonstrate that the framework is capable of constructing objective functions that explicitly account for local operations or for gate design techniques like echo pulses. We demonstrate that the objective function design enables more efficient quantum gate discovery, identifying higher fidelity optimal controls in fewer optimization iterations. 

\section{Methods}

Quantum gates are designed by choosing a set of device controls $\vec{\alpha}$ that generate a time-dependent quantum evolution $U_D(\vec{\alpha},T_c)$ on a quantum device as 
\begin{align}
    U_D(\vec{\alpha},T_{c}) = \mathcal{T}\exp \bigg[-\frac{i}{\hbar}\int_0^{T_c} H_D(\vec{\alpha},\tau) d\tau \bigg],
\end{align}
where $H_D(\vec{\alpha},t)$ is the time-dependent device Hamiltonian, $T_c$ is the total control time, and $\mathcal{T}$ is the time ordering operator. Determining the multiple control parameters $\vec{\alpha}$ that generate a target unitary evolution $U_T$ is typically a significant challenge, and many existing approaches use a local optimization routine like GRAPE, CRAB, or GOAT \cite{khaneja_optimal_2005,caneva_chopped_2011,machnes_tunable_2018}. All approaches require iterative solutions to the Schr\"odinger equation which limits their utility to low-dimensional or short-duration dynamical simulations. Thus minimizing the number of iterations required for these algorithms will greatly reduce the classical computational overhead of gate discovery. We show in this work that, for certain gate discovery applications, engineering the objective function can lead to fewer optimizations to achieve high fidelity device controls.
\par
The typical objective function for optimal control design of unitary gates is the infidelity \cite{werschnik_quantum_2007,glaser_training_2015}
\begin{equation}\label{eq:infidelity}
    \mathcal{G}_0(U_1,U_2) = 1 - \frac{1}{\text{dim($U$)}} \bigg\vert \textrm{Tr}(U^\dagger_1 U_2) \bigg\vert ^2 
\end{equation}
which takes a value between zero and one and is invariant up to a global phase on the gate. However, if the goal of the application is to discover a high-fidelity universal two-qubit quantum gate, the infidelity is incapable of assigning equivalent costs to two universal quantum gates. For example, if $U_1=$ \cnot~and $U_2=$ \cphase~then $\mathcal{G}_0=0.75$. In this case, \cnot~and \cphase~belong to a local equivalence class of universal two-qubit operations \cite{nielsen2002quantum} that are equivalent under specific single-qubit rotations.
\par
A naive approach to addressing this problem is to perform multiple, independent optimizations with different unitary operators of interest defining each instance of the objective function. However, there are generally an uncountable number of possible unitaries that may implement a universal two-qubit entangling gate and, therefore, this approach will require many more solutions of the Schr\"odinger equation. This will dramatically increase the overhead of gate discovery.
\par 
We propose an alternative strategy that defines a functional $\mathcal{F}$ for the device's unitary evolution operator $U_D(\vec{\alpha},T_c)$. This approach recasts the optimization of the infidelity from \eqref{eq:infidelity} into an alternate metric. Notably, the definition of $\mathcal{F}$ is tailored to the application and enables exploration of new gate paradigms. We will present two illustrative examples that 1) account for known high-fidelity single-qubit controls when designing a two-qubit entangling gate, and 2) account for known pulse-echo sequence for implementing two-qubit entangling gates. 
\par
For the first example, we identify a family of control parameters that implement a target two-qubit unitary $U_T$ up to local qubit rotations. Previous strategies to optimize two-qubit unitaries up to local rotations have used Lie theory \cite{watts_optimizing_2015,goerz_optimizing_2015}, but these methods are not easily generalized beyond well understood analytic situations. By contrast, our strategy applies outside of analytic situation by using  optimization formalism that can be solved with numerical methods. We define the functional
\begin{align}\label{eq:F1}
    \mathcal{F}_1(\vec{\theta}, U) &= R(\vec{\theta}_1) U R(\vec{\theta}_2)
\end{align}
where $\vec{\theta} = [\vec{\theta}_1,\vec{\theta}_2]$ refers to a vector of single-qubit rotation angles with
\begin{align}\label{eq:sqrs}
    R(\vec{\theta}) &= R^{(2)}_z(\theta_6)R^{(2)}_y(\theta_5)R^{(2)}_z(\theta_4) \\ \nonumber
    &~~~\otimes R^{(1)}_z(\theta_3)R^{(1)}_y(\theta_2)R^{(1)}_z(\theta_1)
\end{align}
the standard Euler decomposition for qubit rotations in $SU(2)$ on two qubits and $\theta_i$ the $i$th component of $\vec{\theta}$. In \eqref{eq:sqrs}, $R^{(i)}_\mu(\theta) = \exp(-i\theta \sigma^{(i)}_\mu /2 )$ is the standard rotation by an angle $\theta$ around the $\mu$-axis of the Bloch sphere, and $\sigma^{(i)}_\mu$ is the Pauli matrix for direction $\mu~ \in~ \{x,y,z\}$ of qubit $i$.
\par
Using the transformation $\mathcal{F}_1$, we test for equivalence under single-qubit rotations between the target operator $U_1$ and the designed operator $U_2$ via the objective function
\begin{equation}\label{eq:G1}
    \mathcal{G}_1(U_1,U_2) = \min_{\vec{\theta}} \mathcal{G}_0(U_1, \mathcal{F}_1(\vec{\theta}, U_2))
\end{equation}
which can be interpreted as the lowest infidelity obtainable by applying single-qubit rotations before and after $U_2$. Then one can solve an optimal control problem for a target unitary $U_T$
\begin{equation}
    \min_{\vec{\alpha}} \mathcal{G}_1(U_T,U_D(\vec{\alpha},T_c)).
\end{equation}
\par
We remark that the minimization in \eqref{eq:G1} must, in principle, be global in order to saturate the bound of infidelity up to local equivalence of $U_1$ and $U_D(\vec{\alpha},T_c)$. However, we find that in practice an ensemble of local optimizations from different starting points $\vec{\theta}$ is sufficient. Also, \eqref{eq:F1} only requires matrix multiplication to evaluate for a fixed $U$, so the optimization of $\vec{\theta}$ in \eqref{eq:G1} can be performed quite quickly for even moderately sized unitaries.
\par
For the second example, we identify a family of controls that implement a target unitary $U_T$ via an echo-gate interleaved with single-qubit rotations, i.e., we apply the same unitary twice with single-qubit rotations between them. This objective function is motivated by current state-of-the-art two-qubit gates in superconducting systems which are echo pulses that have been used to mitigate noise and reduce gate time \cite{sheldon_procedure_2016}. In this case, we define the functional
\begin{align}\label{eq:F2}
    \mathcal{F}_2({\vec{\theta}},U) &= R(\vec{\theta}_3) U R(\vec{\theta}_{2}) U R(\vec{\theta}_{1})
\end{align}
 with $\vec{\theta} = [\vec{\theta}_1,\vec{\theta}_2,\vec{\theta}_3]$ being the vector of angles for single qubit rotations, before, between, and after the two-qubit evolution. As before, we test for equivalence under $\mathcal{F}_2$ between the target operator $U_1$ and the designed operator $U_2$ via the objective function
\begin{equation}\label{eq:G2}
    \mathcal{G}_2(U_1,U_2) = \min_{\vec{\theta}} \mathcal{G}_0(U_1, \mathcal{F}_2(\vec{\theta},U_2)).
\end{equation}
The corresponding optimal control problem for designing a target unitary $U_T$ is then cast as
\begin{equation}\label{eq:opt_g2}
    \min_{\vec{\alpha}} \mathcal{G}_2(U_T,U_D(\vec{\alpha},T_c)).
\end{equation}
\par 
As before, the minimization in \eqref{eq:G2} must be global to ensure true equivalence, but due to the relative computational efficiency of calculating \eqref{eq:G2}, an ensemble of local optimizations for different $\vec{\theta}$ can be preformed, which we find is sufficient for performing optimal control as in \eqref{eq:opt_g2}.
\par
We next apply these strategies when the target unitary operator is $U_T=$\cnot~and the device of interest is a system of fixed-frequency transmons with microwave-only control \cite{krantz_quantum_2019}. The effective device Hamiltonian for a two-transmon system can be written as
\begin{align}\label{eq:eff_hamiltonian}
    H_D(\vec{\gamma}) &= \omega_{1} \hat{n}_1 + \frac{\delta_1}{2} \hat{n}_1(\hat{n}_1 - 1) \\ \nonumber
    &+ \omega_{2} \hat{n}_{2} + \frac{\delta_{2}}{2} \hat{n}_{2}(\hat{n}_{2} - 1) \\ \nonumber
    &+ J_{1,2} (a^\dagger_1 a_2 + a_1 a^\dagger_2) \\ \nonumber
    &+ \gamma_1(t)(a^\dagger_1+ a_1) \\ \nonumber
    &+ \gamma_2(t)(a^\dagger_2 +a_2),
\end{align}
where $\omega_i,\delta_i, \gamma_i(t)$ are the frequency, anharmonicity, and microwave controls for qubit $i$ and $J_{1,2}$ is always-on coupling between the two transmons \cite{chow_simple_2011,krantz_quantum_2019,sheldon_procedure_2016}. In this model hardware, two-qubit gates are commonly implemented via the cross-resonance effect where one control qubit is driven with a microwave field at the frequency of the second target qubit  \cite{chow_simple_2011}. Optimal control methods have previously been applied to gate design based on the cross-resonance effect, however, those works have used standard objective functions to optimize the pulses \cite{allen_optimal_2017,kirchhoff_optimized_2018}. The main contribution of this work is the development of novel objective functions for gate design, demonstrated in the context of the cross-resonance effect but applicable to any quantum hardware.
\par

\setlength{\tabcolsep}{35pt}
\begin{table}[tbp]
\caption{Hardware parameters are taken from \cite{sheldon_procedure_2016}.}
\begin{center}
    \begin{tabular}{|l|c| }
        \hline \textbf{Parameter} & \textbf{Value (GHz)}  \\ \hline
        $\omega_{1}/2\pi$  &  5.114 \\ \hline
        $\omega_{2}/2\pi$ &  4.914 \\ \hline
        $\delta_1/2\pi$  &  -0.330 \\ \hline
        $\delta_2/2\pi$ &  -0.330 \\ \hline
        $J_{1,2}/2\pi$  &   0.0038 \\ \hline
        $\varepsilon_m/2\pi$  &   0.03 \\ \hline
        $N$ & 22 \\
        \hline
    \end{tabular}
\label{tab:params}
\end{center}
\end{table}
\par 
For our demonstration, we only consider microwave pulses on transmon $1$ ($\gamma_2(t)=0$), which can be used to generate a rotation on transmon 2 dependent on the state of transmon $1$.  We calculate results using these different objective functions with the GOAT method \cite{machnes_tunable_2018}. We expand each control field as a linear combination of $N$ analytic functions, which we choose to be sinusoidal functions
\begin{equation}
    f(\vec{\alpha},t) = \bigg[ \sum_{n}^N \alpha_{n,1}\sin(\alpha_{n,2}t+\alpha_{n,3})\bigg].
\end{equation}
The total microwave control field that is driven in-phase at the dressed frequency of transmon 2 ($\overline{\omega}_2 \approx \omega_2 - J^2/(\omega_1-\omega_2)$) is given by
\begin{align}\label{eq:control_amp}
    \gamma_1(\vec{\alpha},t) = \varepsilon(t)\cos(\overline{\omega}_2t)S(f(\vec{\alpha},t)),
\end{align}
where $S(t)$ is a logistic function that limits the pulse amplitude to a specified range and is defined as
%
%
\begin{equation}
    S(f(\vec{\alpha},t)) = -B - \frac{2B}{1-3\exp(-\frac{g f(\vec{\alpha},t)}{B})}
\end{equation}
where $B=0.08\text{ GHz}/2\pi$ and $g=4$.
\par
Additionally, we have used $\varepsilon(t)$ as a window function to ensure that the control pulses start and stop smoothly at zero. For these demonstrations, the window is chosen to be a flat-top cosine function
\begin{align}
    \varepsilon(t) = 
    \begin{cases} 
      \frac{1-\cos(\pi t/\tau_r)}{2}\varepsilon_m & 0\leq t \leq \tau_r \\
      \varepsilon_m & \tau_r \leq t \leq (\tau_c-\tau_r) \\
      \frac{1-\cos(\pi (\tau_c -t)/\tau_r)}{2} \varepsilon_m & (\tau_c - \tau_r) \leq t \leq \tau_c,
   \end{cases}
\end{align}
where $\tau_c= T_c$ is the total control time, $\varepsilon_m$ scales the magnitude of the control pulse, and $\tau_r$ is the ramp time, which was chosen to be $0.3\tau_c$ to reduce spectral leakage \cite{tripathi_operation_2019}. Parameters for our numerical simulations can be found in Table~\ref{tab:params}. 
\par
We assume that the total pulse time generating two-qubit evolution based on the cross-resonance operation is approximately 200 ns, which has been shown to be an experimentally relevant time scale for this effect \cite{chow_simple_2011,krantz_quantum_2019,tripathi_operation_2019,sheldon_procedure_2016}. In particular, for optimizing $\mathcal{G}_0,\mathcal{G}_1$ we fix $T_c=200$ ns and for the echo-pulse being optimized under $\mathcal{G}_2$, since the pulse is to be applied twice in sequence, we fix $T_c=100$ ns. In all simulations, we consider a smooth microwave pulses, not a sequence of discrete amplitudes, as would be generated in hardware by an arbitrary waveform generator (AWG). This approximation introduces error that is determined by the AWG specifications, and quantifying this error is outside the scope of the present work \cite{motzoi_optimal_2011}. 
\par
In addition to non-ideal effects from AWG sampling, a pressing issue facing superconducting quantum devices are static, non-classical ``$ZZ$'' interactions that emerge in the qubit subspace as a result of the non-computational states of the transmon \cite{ku2020suppression}. Our work takes into account the first three levels of each transmon, partially accounting for this $ZZ$ interaction. However, because we have not considered even higher levels of the transmon, or the effect of a coupling resonator, we do not account for the $ZZ$ interaction completely. Quantifying the residual errors introduced by this approximation will be a focus of future work using more complex models. 
\par
Evaluating the objective functions $\mathcal{G}_1$ and $\mathcal{G}_2$ requires performing an optimization over the set of ancillary parameters $\vec{\theta}$. Thus the optimization of the optimal controls requires performing two optimizations: one over control parameters $\vec{\alpha}$ and one over the ancillary parameters $\vec{\theta}$. However, optimizing both parameter sets via a gradient-based method simultaneously would require solving the Schr\"odinger equation for every update to $\vec{\theta}$, which ignores the fact that optimizing $\vec{\theta}$ for a fixed $U_D(\vec{\alpha},T_c)$ can be performed much more efficiently. In order to take advantage of the dichotomy of computational complexity we implement a heuristic where we iterate between optimizations which are ``easy'' (optimizations of $\vec{\theta}$) and optimizations which are ``hard'' (optimizations of $\vec{\alpha}$).
\par
Specifically, in order to perform the control optimization we break each iteration of the control optimization into two steps. With initial conditions $\vec{\theta}_0,\vec{\alpha}_0$, the first step takes in the initial control guesses $\vec{\alpha}_0$ and optimizes the ancillary parameters which minimize the objective function: $\vec{\theta}_1$. Then we fix $\vec{\theta} = \vec{\theta}_1$ in $\mathcal{F}$ and perform an optimization over $\vec{\alpha}$ to identify a new set of controls $\vec{\alpha}_1$. This is repeated until convergence criteria are reached, or a maximum number of iterations are completed.
\par
We implement the GOAT algorithm using the programming language Julia and various open-source packages. Our implementation uses the Julia package DifferentialEquations.jl to numerically solve the coupled GOAT equations of motion using a order 5/4 Runge-Kutta method with adaptive time stepping \cite{rackauckas2017differentialequations}. We do not implement the rotating wave approximation, which ensures a more accurate estimate of gate fidelity. For the gradient-based control optimization of $\vec{\alpha}$, we use a limited-memory Broyden-Fletcher-Goldfarb-Shanno algorithm with a backtracking line-search method; both of which are implemented in the Optim.jl package \cite{mogensen2018optim}. For the optimization of the ancillary parameters $\vec{\theta}$ we utilize a gradient-free Nelder-Mead algorithm implemented in the Julia package Optim.jl. We limit each optimization to 400 GOAT iterations and define a stopping criteria when the infinity-norm of the gradient falls below 1e-5 or the relative change in the objective function is below 1e-6. For evaluating $\mathcal{G}_1,\mathcal{G}_2$ we perform 5 GOAT iterations before re-optimizing the ancillary parameters $\vec{\theta}$ which requires up to 80 iterations to saturate the 400 GOAT iterations maximum.

\section{Results}

\begin{figure}[tbp]
\centerline{\includegraphics[width=0.45\textwidth]{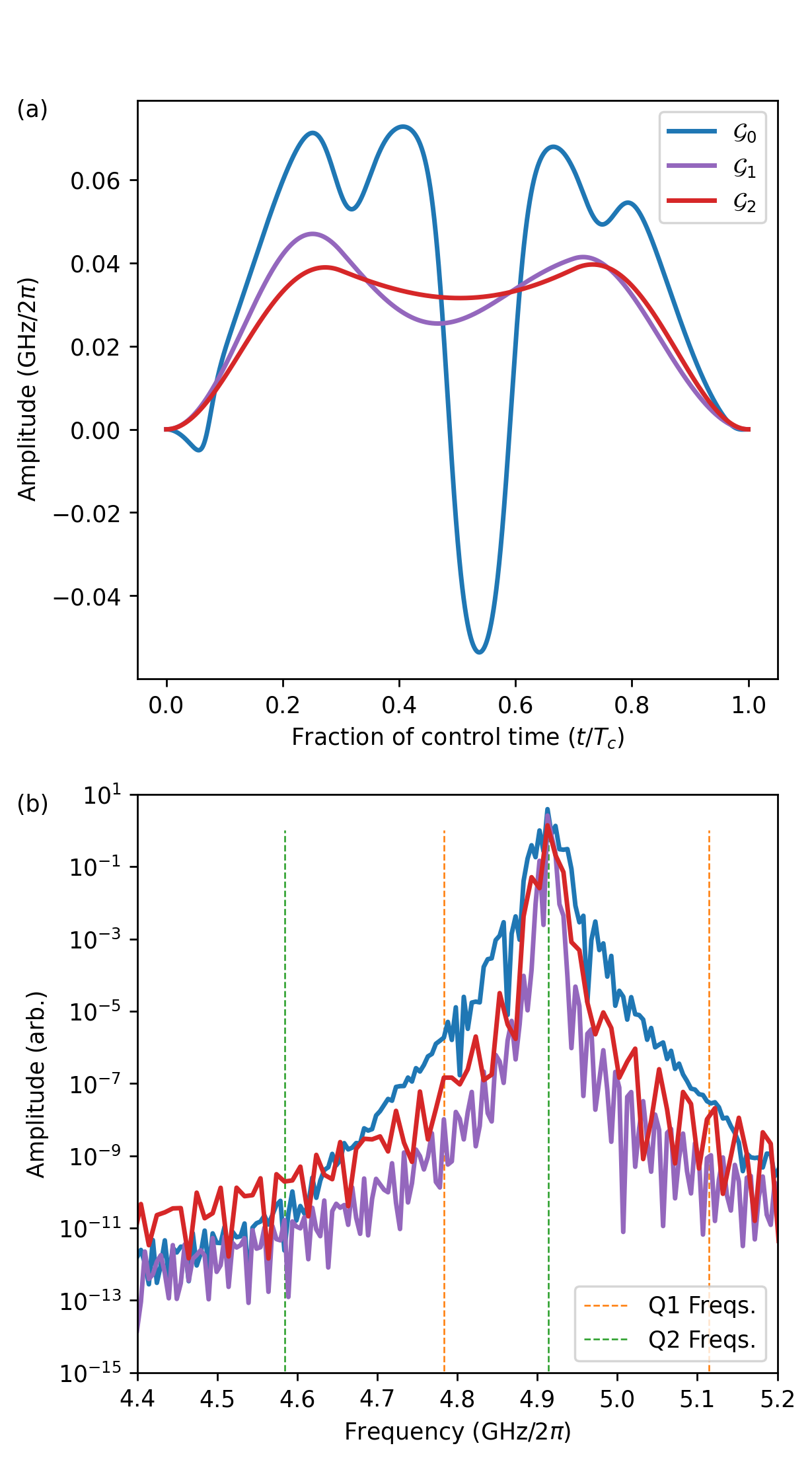}}
\caption{Optimal pulses identified using three different objective functions $\mathcal{G}_0,\mathcal{G}_1,\mathcal{G}_2$ in the time domain (a) and the frequency domain (b). In (a) the the x-axis is labeled in dimensionless units of fractional control time because the optimized pulse durations are 200 ns, 200 ns, and 100 ns for $\mathcal{G}_0,~\mathcal{G}_1$ and $\mathcal{G}_2$, respectively. In (b) we have labeled the lowest energy transitions for the assumed device Hamiltonian, \eqref{eq:eff_hamiltonian}, which includes the transition frequencies for the first transmon (labeled Q1) and second transmon (labeled Q2).}
\label{fig:optimal_pulses}
\end{figure}

To demonstrate that the introduced objective functions $\mathcal{G}_1$ and $\mathcal{G}_2$ lead to more efficient gate discovery than the typical infidelity $\mathcal{G}_0$, we show a set of pulses that have been optimized using each objective function and discuss the characteristics of these optimal pulses, the different convergence behaviour for each objective function, and finally examine the dynamics induced by an optimal pulse found via $\mathcal{G}_2$ on an initial state of interest. 
\par
Using the three different objective functions, we are able to identify a set of pulses in Fig.~\ref{fig:optimal_pulses}a and calculate their spectral density in Fig.~\ref{fig:optimal_pulses}b. In each case, the target unitary to be prepared was a \cnot~gate. It is well known that the cross-resonance effect produces dynamics equivalent to CNOT up to local qubit rotations. We observe that when optimizing $\mathcal{G}_0$ for a 200 ns control time, the control pulse becomes quite complicated as the optimization routine converges to a pulse which implements the CNOT directly.
\par
However, when adding the additional optimization of single qubit rotations into the definition of $\mathcal{G}_1$ via $\mathcal{F}_1$ we observe that the optimal 200 ns pulse is similar to the typical cross-resonance pulse used to generate a \cnot~, i.e. a pulse envelope with constant amplitude \cite{chow_simple_2011,tripathi_operation_2019}, however in this case the pulse differs from the constant-amplitude envelope due to the added effects of higher transmon levels and total pulse time. The similarities are to be expected because the standard cross-resonance pulse is locally equivalent to a \cnot~gate, which is what $\mathcal{G}_1$ is meant to permit \cite{chow_simple_2011,krantz_quantum_2019}.
\par
Beyond the paradigm of locally-equivalent operations, we introduced in \eqref{eq:G2} an objective function intended to identify a 100 ns pulse that, when echoed and interleaved with single qubit rotations, yields a \cnot~gate. Based on the pulse in Fig.~\ref{fig:optimal_pulses}a, we observe that the optimal pulse has low pulse complexity, low amplitude, and high fidelity suggesting that discovering novel gate paradigms can be approached via optimal control and numerical simulation as well as analytic analysis \cite{sheldon_procedure_2016}.

\begin{figure}[tbp]
\centerline{\includegraphics[width=0.45\textwidth]{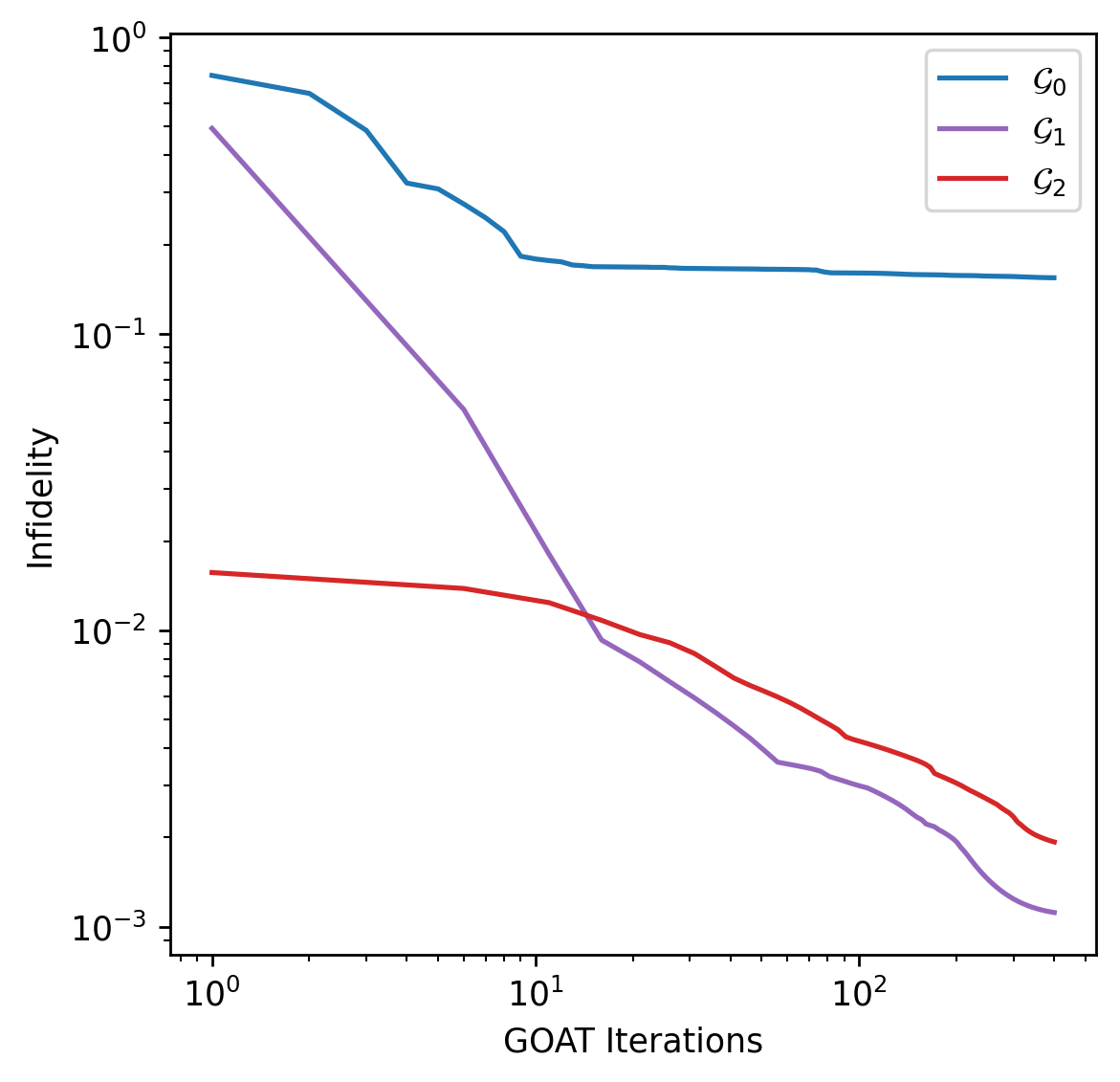}}
\caption{Convergence from an initial guess using three different objective functions $\mathcal{G}_0,\mathcal{G}_1,\mathcal{G}_2$.}
\label{fig:convergence}
\end{figure}

\par 
We next demonstrate that the new objective functions yield a more efficient route to gate discovery. We plot in Fig.~\ref{fig:convergence} the convergence of each objective function with the total number of GOAT iterations. We observe that the new objective functions introduced in this work, $\mathcal{G}_1$ and $\mathcal{G}_2$, converge to lower infidelity solutions in fewer iterations than optimal control using the traditional objective function $\mathcal{G}_0$. This increase in convergence is due to the new objective functions introducing minima in the landscape of possible controls due to the additional degrees of freedoms of single qubit rotations.
\par
\begin{figure}[tbp]
\centerline{\includegraphics[width=0.45\textwidth]{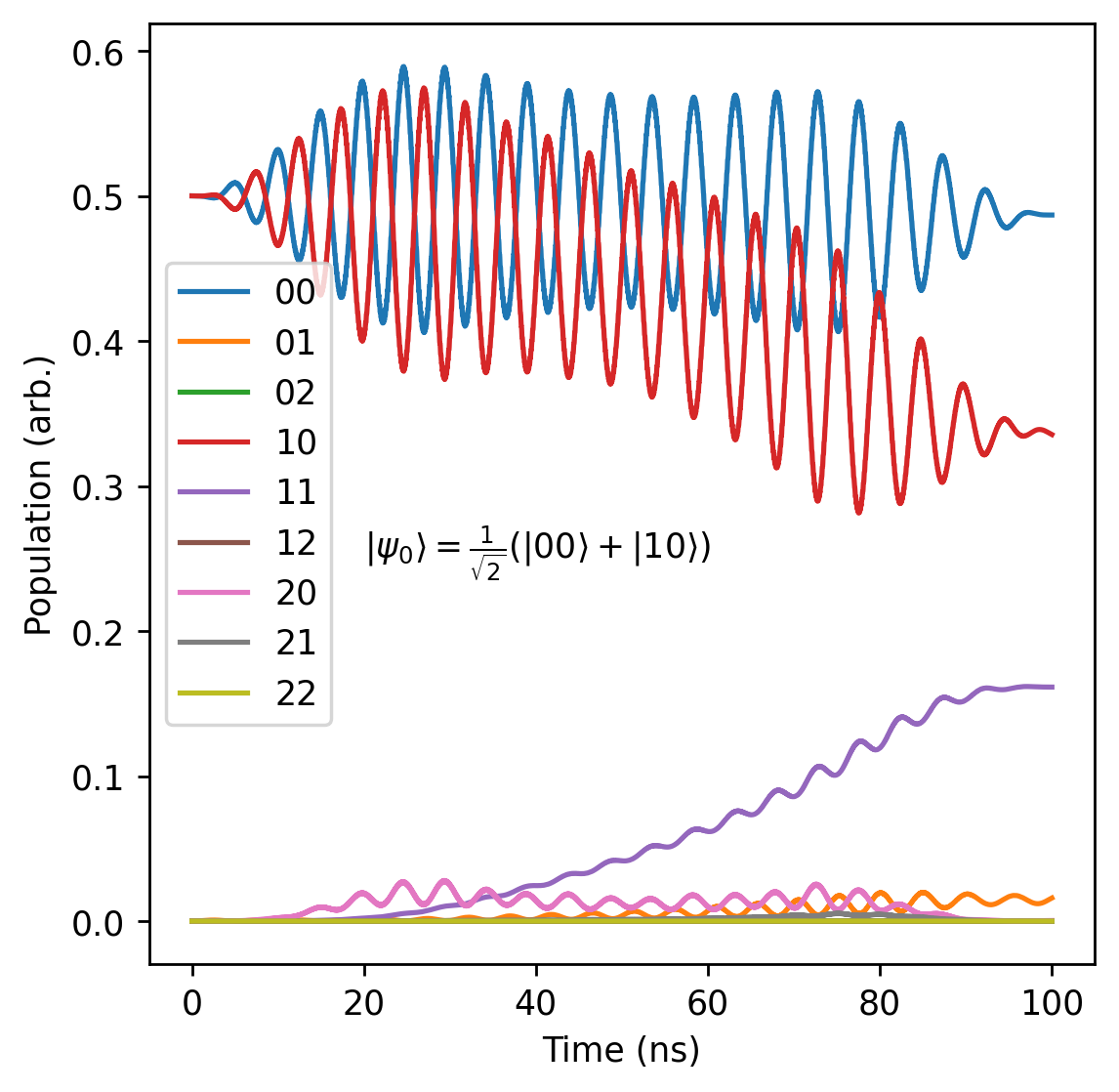}}
\caption{Population dynamics throughout a 100 ns echo pulse found via optimization of $\mathcal{G}_2$, shown in Fig.~\ref{fig:optimal_pulses}. The initial state for the dynamics is $\ket{\psi}_0 = \frac{1}{\sqrt{2}} (\ket{00}+\ket{10})$. The rapid oscillations between the $\ket{10}$ and $\ket{00}$ (and to a lesser extent $\ket{20}$) are due to the off-resonance driving of transmon 1. These dynamics are often neglected through the rotating-wave approximation because they average to zero, but we have chosen to not make the rotating-wave approximation in this work to obtain a more realistic simulation.}
\label{fig:population_dynamics}
\end{figure}
\par
Finally, in order to gain insight into the dynamics being driven by the optimal pulses, we consider in Fig.~\ref{fig:population_dynamics} the state evolution under the pulse found with $\mathcal{G}_2$ in Fig.~\ref{fig:optimal_pulses}a. This optimal pulse should not behave as a \cnot~because it is intended to be echoed and interleaved with single qubit rotations. To verify this visually, we consider the initial state that is given as $\ket{\psi}_0 = \frac{1}{\sqrt{2}} (\ket{00}+\ket{10})$, which is the same state that would be prepared if a Hadamard gate was applied to the control qubit (left-most label) of the product state $\ket{00}$. This is a relevant input state, because if the pulse generated a \cnot~operation, then the final state should be a maximally-entangled Bell state. However, as seen in Fig.~\ref{fig:population_dynamics} this is not the case, which is expected because the single 100 ns pulse was not designed to produce a \cnot.
\par
In addition to the dynamics in Fig.~\ref{fig:population_dynamics} not producing a Bell state, we observe that the dynamics induce some intermediate-time leakage out of the qubit subspace of the control transmon (1). However, we observe that all of the population returns to the qubit subspace at the end of the pulse. This is to be expected because, as can be seen in Fig.~\ref{fig:optimal_pulses}b, the optimal pulse is well localized in the frequency domain, and only perturbatively couples to the higher levels of the transmon.
\section{Conclusions}
We have introduced a way to design objective functions for quantum optimal control problems in quantum logic gate discovery. Our simulations demonstrate that by designing the objective functions with domain and application-specific knowledge leads to optimal controls with lower infidelity and faster convergence than those obtainable with traditional objective functions. Moreover, we have demonstrated that these objective functions can identify optimal controls capable of implementing \cnot~gates in fixed-frequency transmon devices, creating new avenues for gate discovery in superconducting quantum devices.
\par
While the results presented in this work indicate the potential for objective function engineering in quantum gate discovery applications, more detailed studies are needed to fully understand the strengths and weaknesses of this approach. In particular, based on our simulations, we have identified two primary directions that should be explored.
\par
The first direction to be explored should be the impact of pulse ansatz complexity and initial guess on the quality of the optimal solutions. This will require performing an ensemble of optimizations corresponding to independent initial guesses. That study will identify to what degree the results we have observed are typical for the proposed objective functions $\mathcal{G}_1,\mathcal{G}_2$.
\par
The second direction to be explored is the impact of device parameter uncertainty on the pulse fidelity, i.e. quantifying the ability of $\mathcal{G}_1,\mathcal{G}_2$ to identify robust quantum controls. The simulations will require models using a larger number of levels for each transmon in order to determine the impact of leakage and quantify the strength of residual $ZZ$ interactions. Incorporating phenomena such as decoherence, AWG effects, and classical cross-talk should also improve the accuracy of the numerical results and this will be the focus of future work. 
\par
In conclusion, we have proposed a way to engineer objective functions for quantum gate discovery with optimal control theory that is more efficient than using typical objective functions. However, further work is needed in order to understand the strengths and weaknesses of this approach in more realistic applications. To this end, we have discussed potential avenues for future work and believe that our results, although preliminary, open promising routes for quantum gate discovery. 

\section*{Acknowledgment}
This material is based upon work supported by the U.S.~Department of Energy, Office of Science, National Quantum Information Science Research Centers, Quantum Science Center and the U.S.~Department of Energy, Office of Science, Early Career Research Project. This research used resources of the Compute and Data Environment for Science (CADES) at the Oak Ridge National Laboratory, which is supported by the Office of Science of the U.S. Department of Energy under Contract No. DE-AC05-00OR22725.

\bibliographystyle{IEEEtran}
\bibliography{Control,devices}

\end{document}